\documentclass[11pt,preprint,nofootinbib,showpacs,preprintnumbers,amsmath,amssymb]{revtex4}

\usepackage{exscale}
\usepackage{relsize}
\usepackage{graphicx}
\usepackage{dcolumn}
\usepackage{bm}
\usepackage{multirow}
\usepackage{CJK}
\usepackage{diagbox}
\usepackage{subfigure}
\usepackage[T1]{fontenc}
\usepackage{subfigure}

\begin{document}

\begin{CJK*}{GBK}{}

\title{The impact of $a_0^0(980)-f_0(980)$ mixing on the localized $CP$ violations of the $B^-\rightarrow K^- \pi^+\pi^-$ decay}

\author{Jing-Juan Qi \footnote{e-mail: jjqi@mail.bnu.edu.cn}}
\affiliation{\small{Junior College, Zhejiang Wanli University, Zhejiang 315101, China}}

\author{Zhen-Yang Wang \footnote{Corresponding author,e-mail: wangzhenyang@nbu.edu.cn}}
\affiliation{\small{Physics Department, Ningbo University, Zhejiang 315211, China}}

\author{Chao Wang \footnote{e-mail: chaowang@nwpu.edu.cn}}
\affiliation{\small{Center for Ecological and Environmental Sciences, Key Laboratory for Space Bioscience and Biotechnology, Northwestern Polytechnical University, Xi'an 710072, China}}

\author{Zhen-Hua Zhang \footnote{e-mail: zhangzh@usc.edu.cn}}
\affiliation{\small{School of Nuclear and Technology, University of South China, Hengyang, Hunan 421001, China}}

\author{Xin-Heng Guo \footnote{Corresponding author, e-mail: xhguo@bnu.edu.cn}}
\affiliation{\small{College of Nuclear Science and Technology, Beijing Normal University, Beijing 100875, China}}

\date{\today\\}
\begin{abstract}
In the framework of the QCD factorization approach, we study the localized $CP$ violations of the $B^-\rightarrow K^- \pi^+\pi^-$ decay with and without $a_0^0(980)-f_0(980)$ mixing mechanism, respectively, and find that the localized $CP$ violation can be enhanced by this mixing effect when the mass of the $\pi^+\pi^-$ pair is in the vicinity of the $f_0(980)$ resonance. The corresponding theoretical prediction results are $\mathcal{A}_{CP}(B^-\rightarrow K f_0 \rightarrow K^-\pi^+\pi^-)=[0.24, 0.36]$ and $\mathcal{A}_{CP}(B^-\rightarrow K^- f_0(a_0) \rightarrow K^-\pi^+\pi^-)=[0.33, 0.52]$, respectively. Meanwhile, we also calculate the branching fraction of the $B^-\rightarrow K^-f_0(980)\rightarrow K^-\pi^+\pi^-$ decay, which is consistent with the experimental results. We suggest that $a_0^0(980)-f_0(980)$ mixing mechanism should be considered when studying the $CP$ violation of the $B$ or $D$ mesons decays theoretically and experimentally.
\end{abstract}
\pacs{***************}
\maketitle
\end{CJK*}

\section{Introduction}
$CP$ violation plays an important role for the test of the Standard Model (SM) and extractions of the Cabibbo-Kobayashi-Maskawa (CKM) matrix. The processes of nonleptonic decays of $B$ mesons provide us with opportunities for exploring $CP$ violation. In SM, $CP$ violation depends on the weak complex phase in the CKM matrix \cite{Cabibbo:1963yz, Kobayashi:1973fv}. The main uncertainties of $CP$ violation come from the insufficient understanding of strong interaction associated with the nonperturbative QCD. In the past few years, a large amount of experimental data have been collected for $CP$ violation of two body decays of the $B$ meson by $B$ factories, BABAR, Belle, and LHC experiments. The large $CP$ violations have been found by the LHCb Collaboration in the three-body decay channels of $B^\pm\rightarrow \pi^\pm\pi^+\pi^-$ and $B^\pm\rightarrow K^\pm\pi^+\pi^-$ \cite{Aaij:2019jaq,Aaij:2013sfa}. Hence, the exploration of the theoretical mechanism for $CP$ violation becomes interesting in the two- and three-body decays of the $B$ meson.

The nature of the light scalar mesons has attracted much attention for decades since its discovery \cite{Cheng:2005nb,Weinstein:1983gd,Jaffe:1976ig,Kim:2017yur,Abdel-Rehim:2014zwa,Amsler:1995tu,Gorishnii:1983zi}. Because of sharing the same quantum numbers, light scalar mesons play an important role to understand the QCD vacuum. The $a_0^0(980)-f_0(980)$ mixing mechanism has been a hot research topic because of its potential to help understand the structure of scalar mesons. In late 1970s, the $a_0^0(980)-f_0(980)$ mixing effect was first suggested theoretically \cite{Achasov:1979xc}. $a_0^0(980)$ and $f_0(980)$ have
the same spin parity quantum numbers but different isospins. Because of the isospin breaking effect, when they decay into $K\bar{K}$ there exists a difference of 8 $\mathrm{MeV}$ between the charged and neutral kaon thresholds. Up to now, $a_0^0(980)$ and $f_0(980)$ mixing has been studied extensively in various processes and with respect to its different aspects \cite{Colley:1967zz,Achasov:2003se,Kerbikov:2000pu,Krehl:1996rk,Close:2001ay,Kudryavtsev:2002uu,Grishina:2001zj,Black:2002ek,Wu:2010za,Cheng:2020qzc,Aliev:2018bln,Wang:2016wpc,
Sekihara:2014qxa,Liang:2019jtr,Buescher:2003cj,Amsler:2004ps,Hanhart:2003pg,Wang:2004ipa,Roca:2012cv,Close:2015rza,Dorofeev:2011zz}. The signal of this effect was observed for the first time by the BESIII Collaboration in the $J/\psi\rightarrow\phi f_0(980)\rightarrow \phi a_0^0(980)\rightarrow \phi \eta \pi^0$ and $\chi_{c1}\rightarrow a_0^0(980) \pi^0\rightarrow f_0(980)\pi^0\rightarrow \pi^+\pi^-\pi^0$ decays \cite{Ablikim:2018pik}.
Inspired by the fact that $\rho-\omega$ mixing (also due to isospin breaking effect) can induce large $CP$ violations when the invariant mass of the $\pi\pi$ pair is in the $\rho-\omega$ mixing effective area \cite{Guo:1998eg,Wang:2016yrm,Lu:2014oba}, we intend to study the $a_0^0(980)-f_0(980)$ mixing effect on the localized $CP$ violations in three-body decays of the $B$ meson.

In this paper, we will investigate the localized $CP$ violation by $a_0^0(980)-f_0(980)$ mixing and the branching fraction of the $B^-\rightarrow K f_0 \rightarrow K^-\pi^+\pi^-$ decay in the QCDF approach. The remainder of this paper is organized as follows. In Sect. ${\mathrm{\uppercase\expandafter{\romannumeral2}}}$, we present the formalism for $B$ decays in the QCDF approach. In Sect. ${\mathrm{\uppercase\expandafter{\romannumeral3}}}$, we present the $a_0^0(980)-f_0(980)$ mixing mechanism, calculations of the localized $CP$ violation and the branching fraction of the $B^-\rightarrow K f_0 \rightarrow K^-\pi^+\pi^-$ decay. The numerical results are given in Sect. ${\mathrm{\uppercase\expandafter{\romannumeral4}}}$ and we summarize and discuss our work in Sect ${\mathrm{\uppercase\expandafter{\romannumeral5}}}$.

\section{B DECAYS IN THE QCD FACTORIZATION APPROACH}
In the framework of the QCD factorization approach \cite{Beneke:2003zv,Beneke:2001ev}, one can obtain the matrix element $B$ decaying to two mesons $M_1$ and $M_2$ by matching the effective weak Hamiltonian onto a transition operator, which is summarized as follow ($\lambda_p^{(D)}=V_{pb}V_{pD}^*$ with $D=d \ \mathrm{or}\ s$)
\begin{equation}
\langle{M_1M_2}|\mathcal{H}_{eff}|B\rangle=\sum_{p=u,c}\lambda_{p}^{(D)}\langle{M_1M_2}|\mathcal{T}_A^p+\mathcal{T}_B^p|B\rangle,
\end{equation}
where $\mathcal{T}_A^p$ and $\mathcal{T}_B^p$ describe the contributions from non-annihilation and annihilation topology amplitudes, respectively, which can be expressed in terms of the parameters $a_i^p$ and $b_i^p$, respectively, both of which are defined in detail in Ref. \cite{Beneke:2003zv}.

Concretely, $\mathcal{T}_A^p$ contains the contributions from naive factorization, vertex correction, penguin amplitude and spectator scattering and can be expressed as
\begin{equation}\label{a}
\begin{split}
\mathcal{T}_A^p&=\delta_{pu}\alpha_1(M_1M_2)A([\bar{q}_su][\bar{u}D])+\delta_{pu}\alpha_2(M_1M_2)A([\bar{q}_sD][\bar{u}u])\\
&+\alpha_3^p(M_1M_2)\sum_qA([\bar{q}_sD][\bar{q}q])+\alpha_4^p(M_1M_2)\sum_qA([\bar{q}_sq][\bar{q}D])\\
&+\alpha_{3,EW}^p(M_1M_2)\sum_q\frac{3}{2}e_qA([\bar{q}_sD][\bar{q}q])+\alpha_{4,EW}^p(M_1M_2)\sum_q\frac{3}{2}e_qA([\bar{q}_sq][\bar{q}D]),\\
\end{split}
\end{equation}
where the sums extend over $q=u,\,d,\,s$, and $\bar{q}_s(=\bar{u},\, \bar{d}\, \mathrm{or}\,\bar{s})$ denotes the spectator antiquark. The coefficients $\alpha_i^p(M_1M_2)$ and $\alpha_{i,EW}^p(M_1M_2)$ contain all dynamical information and can be expressed in terms of the coefficients $a_i^p$.

As for the power-suppressed annihilation part, we can parameterize it into the following form:
\begin{equation}\label{b}
\begin{split}
\mathcal{T}_B^p&=\delta_{pu}b_1(M_1M_2)\sum_{q'}B([\bar{u}q'][\bar{q}'u][\bar{D}b])+\delta_{pu}b_2(M_1M_2)\sum_{q'}B([\bar{u}q'][\bar{q}'D][\bar{u}b])\\
&+b_3^p(M_1M_2)\sum_{q,q'}B([\bar{q}q'][\bar{q}'D][\bar{q}b])+b_4^p(M_1M_2)\sum_{q,q'}B([\bar{q}q'][\bar{q}'q][\bar{D}b]\\
&+b_{3,EW}^p(M_1M_2)\sum_{q,q'}\frac{3}{2}e_qB([\bar{q}q'][\bar{q}'D][\bar{q}b])+b_{4,EW}^p(M_1M_2)\sum_{q,q'}\frac{3}{2}e_qB([\bar{q}q'][\bar{q}'q][\bar{D}b]),\\
\end{split}
\end{equation}
where $q, q'=u, d, s$ and the sums extend over $q, q'$. The sum over $q'$ arises because a quark-antiquark pair must be created via $g\rightarrow \bar{q}'q'$ after the spectator quark is annihilated.

\section{$a_0^0(980)-f_0(980)$ MIXING MECHANISM, CALCULATION OF CP VIOLATION AND BRANCHING FRACTION}
\subsection{$a_0^0(980)-f_0(980)$ mixing mechanism}
In the condition of turning on the $a_0^0(980)-f_0(980)$ mixing mechanism, we can get the propagator matrix of $a_0^0(980)$ and $f_0(980)$ by summing up all the
contributions of $a_0^0(980)\rightarrow f_0(980)\rightarrow\cdot\cdot\cdot\rightarrow a_0^0(980)$ and $f_0(980)\rightarrow a_0^0(980) \rightarrow\cdot\cdot\cdot\rightarrow f_0(980)$, respectively,  which are expressed as \cite{Sekihara:2014qxa}
\begin{equation}\label{GG}
\begin{split}
\left(
\begin{array}{cccc}
P_{a_0}(s)\quad P_{a_0f_0}(s)\\
 P_{f_0 a_0}(s)\quad P_{f_0}(s)\\
\end{array}
\right)=\frac{1}{D_{f_0}(s)D_{a_0}(s)-|\Lambda(s)|^2}\left(
\begin{array}{cccc}
D_{a_0}(s)\quad \Lambda(s)\\
\Lambda(s)\quad D_{f_0}(s)\\
\end{array}
\right),
\end{split}
\end{equation}
where $P_{a_0}(s)$ and $P_{f_0}(s)$ are the propagators of $a_0$ and $f_0$, respectively, $P_{a_0f_0}(s)$, $P_{f_0 a_0}(s)$ and $\Lambda(s)$ arise due to the $a_0^0(980)-f_0(980)$ mixing effect, and $D_{a_0}(s)$ and $D_{f_0}(s)$ are the denominators for the propagators of $a_0$ and $f_0$ when the $a_0^0(980)-f_0(980)$ mixing effect is  absent, respectively, which can be expressed as follows in the Flatt$\acute{\mathrm{e}}$ parametrization:
\begin{equation}\label{DaDf}
\begin{split}
D_{a_0}(s)&=m_{a_0}^2-s-i\sqrt{s}[\Gamma_{\eta\pi}^{a_0}(s)+\Gamma_{K\bar{K}}^{a_0}(s)],\\
D_{f_0}(s)&=m_{f_0}^2-s-i\sqrt{s}[\Gamma_{\pi\pi}^{f_0}(s)+\Gamma_{K\bar{K}}^{f_0}(s)],\\
\end{split}
\end{equation}
where $m_{a_0}$ and $m_{f_0}$ are the masses of the $a_0$ and $f_0$ mesons, with the decay width $\Gamma^a_{bc}$ can being presented as

\begin{equation}\label{Tabc}
\begin{split}
\Gamma_{bc}^a(s)=\frac{g_{abc}^2}{16\pi\sqrt{s}}\rho_{bc}(s)\quad \mathrm{with} \quad \rho_{bc}(s)=\sqrt{[1-\frac{(m_b-m_c)^2}{s}][1+\frac{(m_b-m_c)^2}{s}]}.\\
\end{split}
\end{equation}

It was pointed out that the contribution from the amplitude of $a_0^0(980)-f_0(980)$ mixing is convergent and can be written as an expansion in the $K\bar{K}$ phase space when only $K\bar{K}$ loop contributions are considered \cite{Achasov:1979xc,Achasov:2017zhu},
\begin{equation}\label{a0f0}
\begin{split}
\Lambda(s)_{K\bar{K}}&=\frac{g_{a_0K^+K^-}g_{f_0K^+K^-}}{16\pi}\bigg\{i\bigg[\rho_{K^+K^-}(s)-\rho_{K^0\bar{K}^0}(s)\bigg]
-\mathcal{O}(\rho_{K^+K^-}^2(s)-\rho_{K^0\bar{K}^0}^2(s))\bigg\},\\
\end{split}
\end{equation}
where $g_{a_0K^+K^-}$ and $g_{f_0K^+K^-}$ are the effective coupling constants. Since the mixing mainly comes from the $K\bar{K}$ loops,
we can adopt $\Lambda (s)\approx \Lambda_{K\bar{K}}(s)$.

\subsection{Decay amplitudes, localizd CP violation and branching fraction}
\begin{figure}[ht]
\centerline{\includegraphics[width=0.6\textwidth]{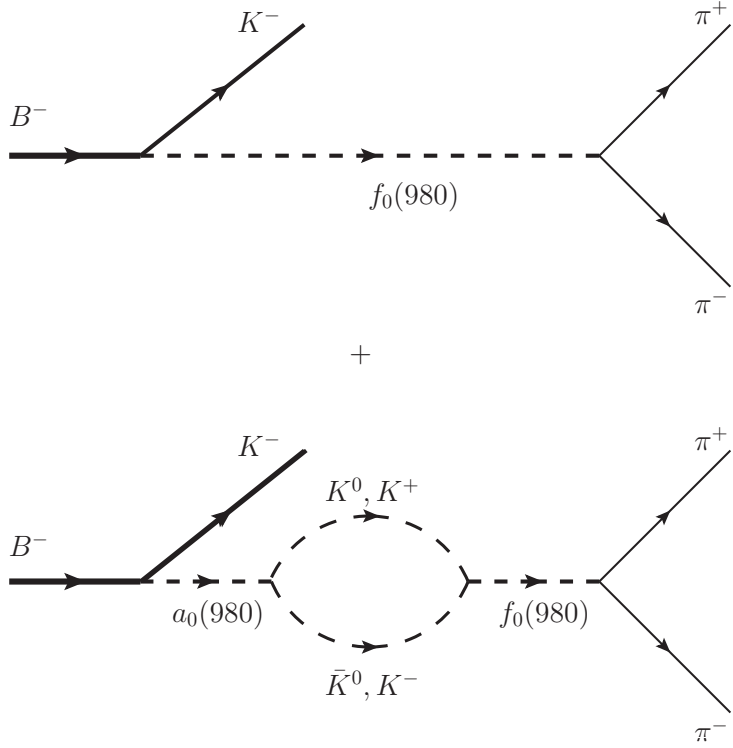}}
\caption{The Feynman diagram for the $B^-\rightarrow K^-\pi^+\pi^-$ decay with the $a_0^0(980)-f_0(980)$ mixing mechanism.}
\label{Mixing}
\end{figure}

With the $a_0^0(980)-f_0(980)$ mixing being considered, the process of the $B^-\rightarrow K^-\pi^+\pi^-$ decay is shown in Fig. \ref{Mixing} and the amplitude can be expressed as

\begin{equation}\label{Av}
\begin{split}
\mathcal{M}&=\langle K^-\pi^+\pi^-|\mathcal{H}^T|B^-\rangle+\langle K^-\pi^+\pi^-|\mathcal{H}^P|B^-\rangle,\\
\end{split}
\end{equation}
in which $\mathcal{H}^T$ and $\mathcal{H}^P$ are the tree and penguin operators, respectively, and we have
\begin{equation}\label{TPAm}
\begin{split}
\langle K^-\pi^+\pi^-|\mathcal{H}^T|B^-\rangle &=\frac{g_{f_0\pi\pi}T_{f_0}}{D_{f_0}}+\frac{g_{f_0\pi\pi}T_{a_0}\Lambda}{D_{a_0}D_{f_0}-\Lambda^2},\\
\langle K^-\pi^+\pi^-|\mathcal{H}^P|B^-\rangle&=\frac{g_{f_0\pi\pi}P_{f_0}}{D_{f_0}}+\frac{g_{f_0\pi\pi}P_{a_0}\Lambda}{D_{a_0}D_{f_0}-\Lambda^2},\\
\end{split}
\end{equation}
where $T_{a_0(f_0)}$ and $P_{a_0(f_0)}$ represent the tree and penguin diagram amplitudes for $B\rightarrow K a_0(f_0)$ decay, respectively. Substituting Eq. (\ref{TPAm}) into Eq. (\ref{Av}), the total amplitude of the decay $B^-\rightarrow K^- f_0(a_0) \rightarrow K^-\pi^+\pi^-$ can be written as
\begin{equation}\label{M3}
\begin{split}
\mathcal{M}(B^-\rightarrow K^-\pi^+\pi^-)=\frac{g_{f_0\pi\pi}}{D_{f_0}}\mathcal{M}(B^-\rightarrow K^-f_0)
+\frac{g_{f_0\pi\pi}\Lambda}{D_{a_0}D_{f_0}-\Lambda^2}\mathcal{M}(B^-\rightarrow K^-a_0).\\
\end{split}
\end{equation}

In the QCD factorization approach, we derive the amplitudes of the $B^-\rightarrow K^-f_0$ and $B^-\rightarrow K^-a_0$ decays, which are
\begin{equation}\label{AR1}
\begin{split}
\mathcal{M}(B^-\rightarrow K^-f_0)&=-\frac{G_F}{\sqrt{2}}\sum_{p=u,c}\lambda_p^{(s)}\bigg\{(\delta_{pu}a_1+a_4^p-r_\chi^K a_6^p+a_{10}^p-r_\chi^K a_8^p)_{f_0^uK}(m_B^2-m_{f_0}^2)f_KF_0^{Bf_0^u}(m_K^2)\\
&-(\delta_{pu}a_2+2a_3^p+2a_5^p+\frac{1}{2}a_9^p+\frac{1}{2}a_7^p)_{Kf_0^u}(m_B^2-m_{K}^2)\bar{f}_{f_0^u}F_0^{BK}(m_{f_0}^2)\\
&-(a_3^p+a_5^p+a_4^p-r_\chi^fa_6^p-\frac{1}{2}a_9^p-\frac{1}{2}a_7^p-\frac{1}{2}a_{10}^p+\frac{1}{2}r_\chi^fa_8^p)_{Kf_0^s}(m_B^2-m_K^2)\bar{f}_{f_0^s}F_0^{BK}(m_{f_0}^2)\\
&+(\delta_{pu}b_2+b_3^p+b_{3,EW})_{Kf_0^u}f_B\bar{f}_{f_0^u}f_K+(\delta_{p,u}b_2+b_3^p-\frac{1}{2}b_{3,EW})_{Kf_0^s}f_B\bar{f}_{f_0^s}f_K\bigg\},\\
\end{split}
\end{equation}
and
\begin{equation}\label{AR2}
\begin{split}
\mathcal{M}(B^-\rightarrow K^-a_0)&=-\frac{G_F}{\sqrt{2}}\sum_{p=u,c}\lambda_p^{(s)}\bigg\{(\delta_{pu}a_1+a_4^p-r_\chi^K a_6^p+a_{10}^p-r_\chi^K a_8^p)_{a_0K}(m_B^2-m_{a_0}^2)F_0^{Ba_0}(m_K^2)f_K\\
&-(\delta_{pu}a_2+\frac{3}{2}a_9^p+\frac{3}{2}a_7^p)_{Ka_0}(m_B^2-m_K^2)F_0^{B\rightarrow K}(m_{a_0}^2)f_{a_0}\\
&+(\delta_{pu}b_2+b_3^p+b_{3,EW}^p)_{a_0K}f_Bf_{a_0}f_K\bigg\},\\
\end{split}
\end{equation}
respectively, where $G_F$ represents the Fermi constant, $f_B$, $f_K$, $\bar{f}_{f_0}$ and $f_{a_0}$ are the decay constants of the $B$, $K$, $f_0$, and $a_0$, $\bar{f}_s=f_s\frac{2m_\pi^2}{m_b(\mu)(m_u(\mu)+m_s(\mu))}$ (where $\mu$ is the scale parameter), $F_0^{Bf_0^u}(m_K^2)$, $F_0^{BK}(m_{f_0}^2)$ and $F_0^{Ba_0}(m_K^2)$ are the form factors for the $B$ to $f_0$, $K$ and $a_0$ transitions, respectively.

By integrating the numerator and denominator of the differential $CP$ asymmetry parameter, one can obtain the localized integrated $CP$ asymmetry,  which can be measured by experiments and takes the following form in the region $R$:
\begin{equation}\label{ACP}
\begin{split}
A_{CP}^R=\frac{\int_Rdsds'(\mid\mathcal{M}\mid^2-\mid\mathcal{\bar{M}}\mid^2)}{\int_Rdsds'(\mid\mathcal{M}\mid^2+\mid\mathcal{\bar{M}}\mid^2)},\\
\end{split}
\end{equation}
where $s$ and $s'$ are the invariant masses squared of $\pi\pi$ or $K\pi$ pair in our case, and $\mathcal{\bar{M}}$ is the decay amplitude of the $CP$-conjugate process.

Since the decay process $B^-\rightarrow K^-\pi^+\pi^-$ has a three-body final state, the branching fraction of this decay can be expressed as \cite{Agashe:2014kda}
\begin{equation}\label{BRBR}
\begin{split}
\mathcal{B}=\frac{\tau_B}{(2\pi)^5 16m_B^2}\int ds |\mathbf{p}_1^*||\mathbf{p}_3|\int d\Omega_1^* \int d\Omega_3 |\mathcal{M}|,\\
\end{split}
\end{equation}
in which $\Omega_1^*$  and $\Omega_3$ are the solid angles for the final $\pi$ in the $\pi\pi$ rest frame and for the final $K$ in the $B$ meson rest frame, respectively, $|\mathbf{p}_1^*|$ and $|\mathbf{p}_3|$  are the norms of the three-momenta of final-state $\pi$ in the $\pi\pi$ rest frame, and $K$ in the $B$ rest frame, respectively, which take the following forms:

\begin{equation}\label{PP}
\begin{split}
|\mathbf{p}_1^*|&=\frac{\sqrt{\lambda(s,m_\pi^2,m_\pi^2)}}{2\sqrt{s}},\\
|\mathbf{p}_3|&=\frac{\sqrt{\lambda(m_B^2,m_K^2,s)}}{2m_B},\\
\end{split}
\end{equation}
where $\lambda(a,b,c)$ is the K$\ddot{a}$ll$\acute{\mathrm{e}}$n function and with the form $\lambda(a,b,c)=a^2+b^2+c^2-2(ab+ac+bc)$.
\section{Numerical results}
\begin{figure}[htbp]
    \centering
    \subfigure[]{
        \includegraphics[width=3in]{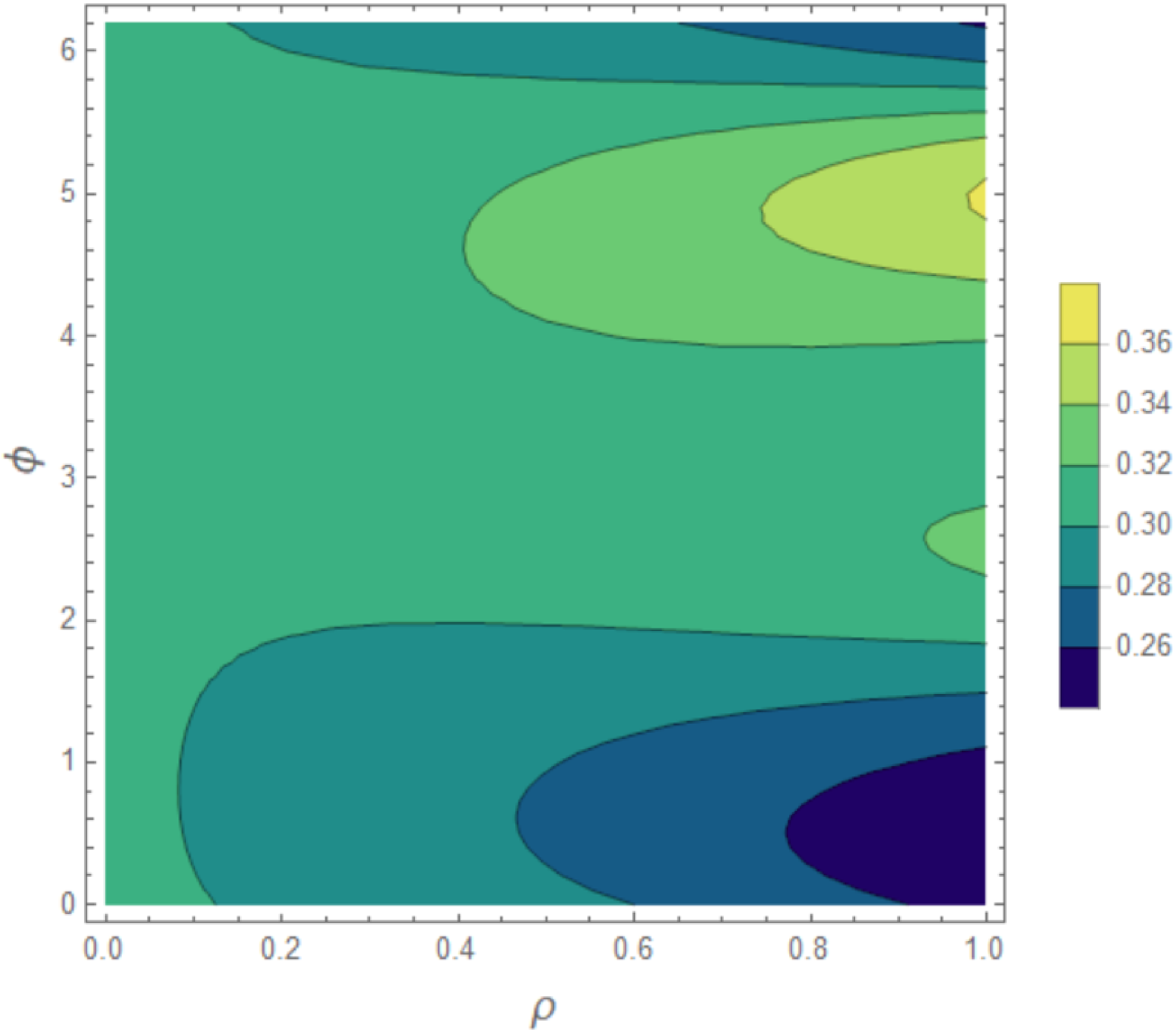}}
    \subfigure[]{
        \includegraphics[width=2.98in]{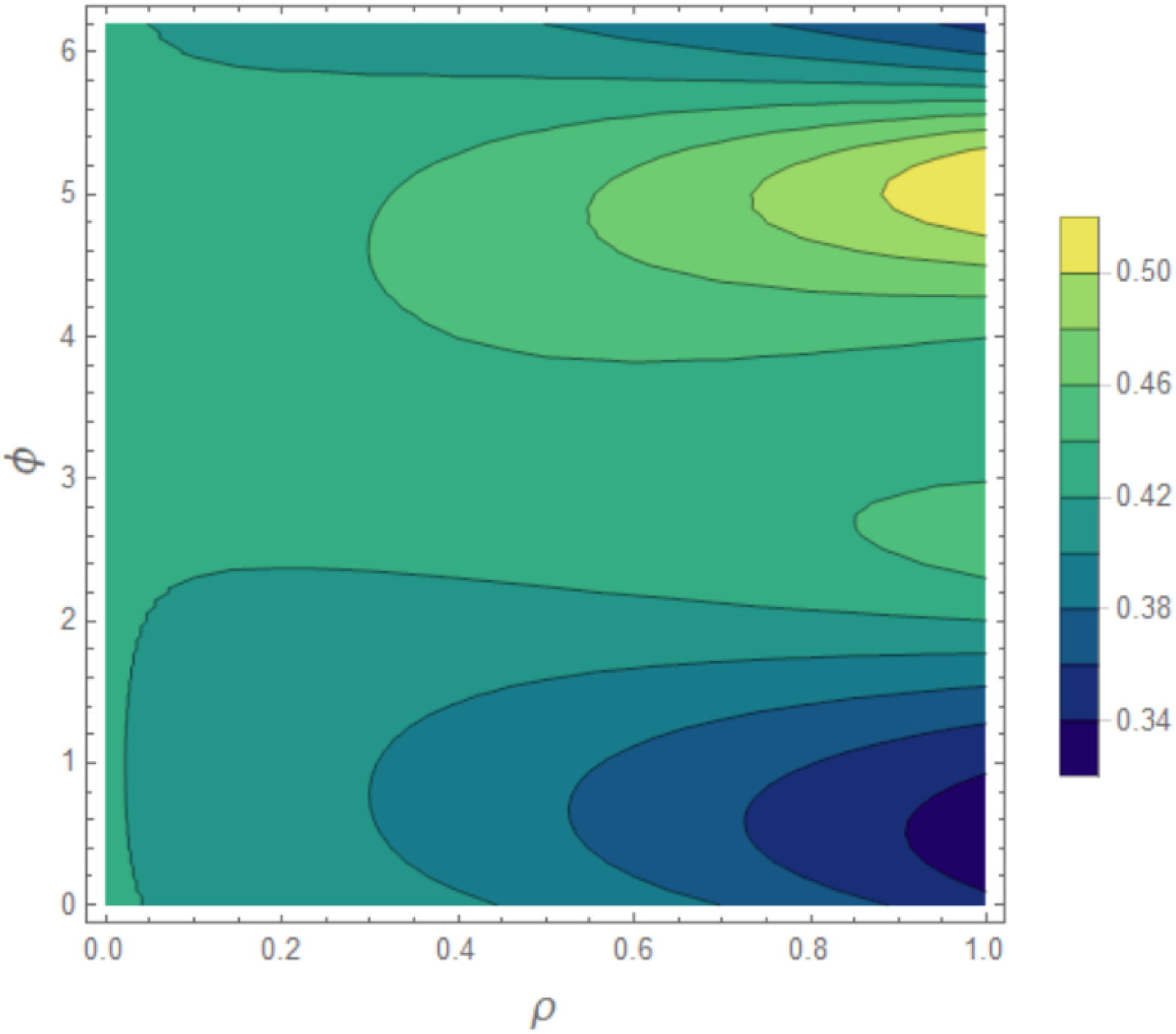}}
    \caption{The localized $CP$ violation of the $B^-\rightarrow K^-f_0 \rightarrow K^-\pi^+\pi^-$ decay (a) without the $a_0^0(980)-f_0(980)$ mixing mechanism,  (b) with the $a_0^0(980)-f_0(980)$ mixing mechanism.}
    \label{ACPFIG}
\end{figure}

\begin{figure}[ht]
\centerline{\includegraphics[width=3.3in]{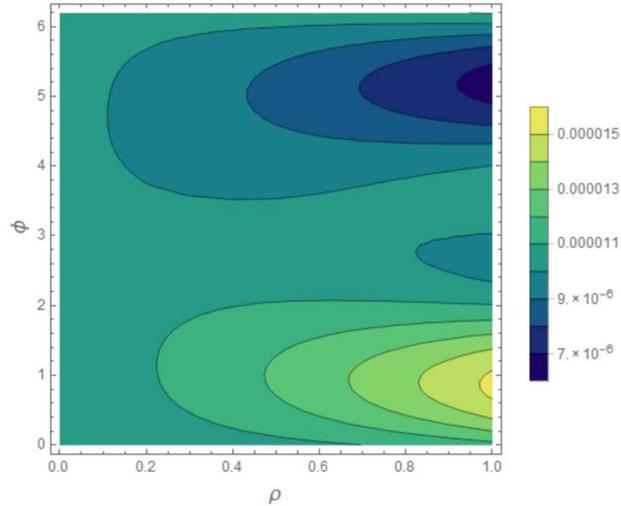}}
\caption{The branching fraction of the $B^-\rightarrow K^-f_0 \rightarrow K^-\pi^+\pi^-$ decay.}
\label{BR}
\end{figure}

When dealing with the contributions from the hard spectator and the weak annihilation, we encounter the singularity problem of infrared divergence $X=\int_0^1dx/(1-x)$. One can adopt the method in Refs. \cite{Beneke:2003zv,Beneke:2001ev,Cheng:2005nb} to parameterize the endpoint divergence as $X_{H,A}=(1+\rho_{H,A} e^{i\phi_{H,A}})\ln\frac{m_B}{\Lambda_h}$, with $\Lambda_h$ being a typical scale of order 0.5 $\mathrm{GeV}$, $\rho_{H,A}$ an unknown real parameter and $\phi_{H,A}$ the free strong phase in the range $[0,2\pi]$. For convenience, we use the notations $\rho=\rho_{H,A}$ and $\phi=\phi_{H,A}$. In  our calculations, we adopt $\rho\in[0, 1]$ and $\phi\in[0, 2\pi]$ for the two-body $B^- \rightarrow K^-f_0$ and $B^- \rightarrow K^-a_0$ decays. The first term of Eq. (\ref{M3}) is the amplitude of the $B^-\rightarrow K^-\pi^+\pi^-$ decay without the effect of the $a_0^0(980)-f_0(980)$ mixing when the mass of the $\pi^+\pi^-$ pair is in the vicinity of the $f_0(980)$ resonance. Substituting this term into Eq. (\ref{ACP}), we can get the localized $CP$ violation of the $B^-\rightarrow K^- f_0 \rightarrow K^-\pi^+\pi^-$ decay when
we take the the integration interval as $[m_{f_0}-\Gamma_{f_0}, m_{f_0}+\Gamma_{f_0}]$, which is $\mathcal{A}_{CP}(B^-\rightarrow K f_0 \rightarrow K^-\pi^+\pi^-)=[0.24, 0.36]$ and shown in Fig. {\ref{ACPFIG}} (a). Substituting Eqs. (\ref{AR1}) and (\ref{AR2}) into Eq. (\ref{M3}), one can also get the total amplitude of the $B^-\rightarrow K^- f_0(a_0)\rightarrow K^-\pi^+\pi^-$ decay with the $a_0^0(980)-f_0(980)$ mixing mechanism. Then inserting it into Eq. (\ref{ACP}), we can also get the result of the localized $CP$ violation in the presence of $a_0^0(980)-f_0(980)$ mixing by integrating the same integration interval as above.  The predicted result is $\mathcal{A}_{CP}(B^-\rightarrow K^- f_0(a_0) \rightarrow K^-\pi^+\pi^-)=[0.33, 0.52]$, which is plotted in Fig. {\ref{ACPFIG}} (b). Obviously, the $CP$ violating asymmetry in Fig. {\ref{ACPFIG}} (b) is significantly larger than that in Fig. {\ref{ACPFIG}} (a). Thus, we conclude that the $a_0^0(980)-f_0(980)$ mixing mechanism can induce larger localized $CP$ violation for the $B^-\rightarrow K^-\pi^+\pi^-$ decay. However, compared with the contribution from first term in Eq. ({\ref{M3}}), that from the second term is very small and even can be ignored when calculating the branching fraction, thus we have $\mathcal{B}(B^-\rightarrow K^-f_0(a_0)  \rightarrow K^-\pi^+\pi^-)\approx \mathcal{B}(B^-\rightarrow K f_0 \rightarrow K^-\pi^+\pi^-)$. Then, we calculate the branching fraction of the $B^-\rightarrow  K f_0 \rightarrow K^-\pi^+\pi^-$ decay combining the first term in Eq. ({\ref{M3}}), Eqs. (\ref{AR1}) and (\ref{BRBR}), the theoretical result is $\mathcal{B}(B^-\rightarrow K^- f_0 \rightarrow K^-\pi^+\pi^-)=[6.50, 15.0]\times10^{-6}$ which is plotted in Fig. {\ref{BR}}. This result is consistent with the experimental result $\mathcal{B}(B^-\rightarrow K f_0 \rightarrow K^-\pi^+\pi^-)=(9.4^{+1.0}_{-1.2})\times10^{-6}$ \cite{Tanabashi:2018oca} when the divergence parameter ranges are taken as $\rho\in[0, 1]$ and $\phi\in[0, 2\pi]$.

\section{SUMMARY AND DISCUSSION}
In this work, we studied the localized integrated $CP$ violation of the $B^-\rightarrow K^-f_0(a_0) \rightarrow K^-\pi^+\pi^-$ decay considering the $a_0^0(980)-f_0(980)$ mixing mechanism in the QCD factorization approach. We found the localized integrated $CP$ violation is enlarged due to the $a_0^0(980)-f_0(980)$ mixing effect. Without the $a_0^0(980)-f_0(980)$ mixing, the localized $CP$ violation was found to be $\mathcal{A}_{CP}(B^-\rightarrow K f_0 \rightarrow K^-\pi^+\pi^-)=[0.24, 0.36]$, while $\mathcal{A}_{CP}(B^-\rightarrow K^- f_0(a_0) \rightarrow K^-\pi^+\pi^-)=[0.33, 0.52]$ when this mixing effect is considered.  In addition, we also calculated the branching fraction of the $B^-\rightarrow K^-f_0 \rightarrow K^-\pi^+\pi^-$ decay, and obtained $\mathcal{B}(B^-\rightarrow K f_0 \rightarrow K^-\pi^+\pi^-)=[6.50, 15.0]\times10^{-6}$ as shown in Fig. {\ref{BR}}, which agrees the experimental result $\mathcal{B}(B^-\rightarrow K f_0 \rightarrow K^-\pi^+\pi^-)=9.4^{+1.0}_{-1.2}\times 10^{-6}$ well. Since the mixing term is very small, while calculating the branching fraction we can take the approximation $\mathcal{B}(B^-\rightarrow K^- f_0(a_0)\rightarrow K^-\pi^+\pi^-)\approx \mathcal{B}(B^-\rightarrow K f_0 \rightarrow K^-\pi^+\pi^-)$ by ignoring the $a_0^0(980)-f_0(980)$ mixing effect. However, for $CP$ violation, this mixing effect does contribution a lot and cannot be neglected. The same situation is also expended for other $B$ or $D$ mesons decay channels. We thus suggest that $a_0^0(980)-f_0(980)$ mixing mechanism should be considered when studying the heavy meson decays both theoretically and experimentally when this mixing effect could exist.

\begin{appendix}
\section{THEORETICAL INPUT PARAMETERS}
In the numerical calculations, we should input distribution amplitudes and the CKM matrix elements in the Wolfenstein parametrization. For the CKM matrix elements,
which are determined from experiments, we use
the results in Ref. \cite{Agashe:2014kda}:
\begin{equation}\label{CKM}
\begin{split}
&\bar{\rho}=0.117\pm0.021, \quad \bar{\eta}=0.353\pm0.013, \\
&\lambda=0.225\pm0.00061,\quad A=0.811^{+0.023}_{-0.024},\\
\end{split}
\end{equation}
where
\begin{equation}\label{rhoyita}
\begin{split}
\bar{\rho}=\rho(1-\frac{\lambda^2}{2}), \quad \bar{\eta}=\eta(1-\frac{\lambda^2}{2}). \\
\end{split}
\end{equation}

The effective Wilson coefficients used in our calculations are taken from Ref. \cite{Wang:2014hba}:
\begin{equation}\label{C}
\begin{split}
&c'_1=-0.3125, \quad c'_2=1.1502, \\
&c'_3=2.433\times10^{-2}+1.543\times10^{-3}i,\quad c'_4=-5.808\times10^{-2}-4.628\times10^{-3}i, \\
&c'_5=1.733\times10^{-2}+1.543\times10^{-3}i,\quad c'_6=-6.668\times10^{-2}-4.628\times10^{-3}i, \\
&c'_7=-1.435\times10^{-4}-2.963\times10^{-5}i,\quad c'_8=3.839\times10^{-4}, \\
&c'_9=-1.023\times10^{-2}-2.963\times10^{-5}i,\quad c'_{10}=1.959\times10^{-3}. \\
\end{split}
\end{equation}

For the masses appeared in $B$ decays, we use the following values \cite{Agashe:2014kda} (in units of $\mathrm{GeV}$):
\begin{equation}
\begin{split}
m_u&=m_d=0.0035,\quad m_s=0.119, \quad m_b=4.2,\quad m_q=\frac{m_u+m_d}{2},\quad m_{\pi^\pm}=0.14,\\
m_{B^-}&=5.279,\quad m_{K^-}=0.494,\quad m_{f_0(980)}=0.990,\quad m_{a_0^0(980)}=0.980,\\
\end{split}
\end{equation}
while for the widthes we use (in units of $\mathrm{GeV}$) \cite{Agashe:2014kda}
\begin{equation}
\Gamma_{f_0(980)}=0.074,\quad\Gamma_{a_0^0(980)}=0.092.\\
\end{equation}

The following numerical values for the decay constants are used \cite{Cheng:2013dua,Cheng:2010yd,Cheng:2005nb} (in units of $\mathrm{GeV}$):
\begin{equation}
\begin{split}
f_{\pi^\pm}&=0.131,\quad f_{B^-}=0.21\pm0.02, \quad f_{K^-}=0.156\pm0.007,  \\
 \bar{f}_{f_0(980)}&=0.370\pm0.02, \quad\bar{f}_{a_0^0(980)}=0.365\pm0.02.\\
\end{split}
\end{equation}

As for the form factors, we use \cite{Cheng:2005nb}
\begin{equation}
\begin{split}
F_0^{B\rightarrow K}(0)&=0.35\pm0.04,\quad F_0^{B\rightarrow f_0(980)}(0)=0.25, \quad F_0^{B\rightarrow a_0^0(980)}(0)=0.25.\\
\end{split}
\end{equation}

The values of Gegenbauer moments at $\mu=1 \mathrm{GeV}$ are taken from \cite{Cheng:2005nb}:
\begin{equation}
\begin{split}
B_{1,f_0(980)}&=-0.78\pm0.08,\quad B_{3,f_0(980)}=0.02\pm0.07,\\
 B_{1,a_0^0(980)}&=-0.93\pm0.10,\quad B_{3,a_0^0(980)}=0.14\pm0.08.\\
\end{split}
\end{equation}
\end{appendix}

\acknowledgments
This work was supported by National Natural Science Foundation of China (Projects Nos. 11775024, 11705081, 11805153, 11947001 and 11605150), Natural Science Foundation of Zhejiang Province (No. LQ21A050005) and the Fundamental Research Funds for the Provincial Universities of Zhejiang Province.

\end{document}